\documentclass[11pt]{article}
\usepackage{moriond,epsfig,amssymb}

\def\Journal#1#2#3#4{{#1} {\bf #2}, #3 (#4)}

\def\PLB{{\em Phys. Lett.}  B}

\def\PRD{{\em Phys. Rev.} D}

\def\RMP{{\em Rev. Mod. Phys.}}
\def\JHEP{{\em JHEP}}
\def\NPPS{{\em Nucl. Phys. Proc. Suppl} }

\def\be{\begin{equation}}
\def\ee{\end{equation}}
\def\bea{\begin{eqnarray}}
\def\eea{\end{eqnarray}}

\begin{document}
\vspace*{4cm}
\title{Instanton traces in lattice gluon correlation functions}

\author{ Ph. Boucaud$^1$, F. de Soto$^{2,3}$, J.P. Leroy$^1$, A. Le Yaouanc$^1$, J. Micheli$^1$, O. Pene$^1$, 
J. Rodriguez--Quintero$^2$}

\address{(1) Laboratoire de Physique Theorique, 
Universit\'e Paris XI, 91405 Orsay Cedex (FRANCE)
(2) Departamento de Fisica Aplicada. Universidad de Huelva
Avda. Fuerzas Armadas s/n. 21000 Huelva (SPAIN) 
(3) Laboratoire de Physique Subatomique et Cosmologie. 
53 av. des Martyrs. 38026 Grenoble (FRANCE)}

\maketitle\abstracts{
Strong coupling constant computed in Landau gauge and MOM renormalization scheme from lattice two
and three gluon Green Functions exhibits an unexpected behavior in the deep IR, showing a maximum
value around $1 {\rm GeV}$. We analise this coupling below this maximum within a semiclassical
approach, were gluon degrees of freedom at very low energies are  described in terms of the
classical solutions of the lagrangian, namely instantons. We provide some new results concerning
the relationship between instantons and the low energy dynamics of QCD, by analising gluon two-
and three-point Green functions separately and with the help of a cooling procedure to eliminate
short range correlations.}

The running of the strong coupling constant is one of the first predictions that one can extract
from perturbation theory in QCD and its behavior at high energies (where perturbation theory
makes sense) has been measured in several experiments~\cite{pdg}. At lower energies, non
perturbative effects appear, and some other alternative to perturbation theory has to be used.
In this note we will report on the analysis at the light of semiclassical approximations of
numerical results about the low-energy QCD coupling obtained by the use of lattice calculations.
The IR behavior of the coupling constant, being a non physically measurable quantity,  is
nevertheless an interesting object in order to get a better understanding of the non 
perturbative nature of the theory, as one can compare the results from lattice simulations with
some phenomenological model, gaining insight on the phenomena happening at low energies  on QCD.
In this sense, this paper will recall the lattice calculation of the coupling constant from two-
and three-point gluon Green functions, and will present some evidences concerning the influence
of semiclassical solutions of the theory (in particular, instantons) over low energy dynamics.

\section{QCD coupling constant}

\begin{figure}
\begin{center}
\begin{tabular}{c c}
\includegraphics[width=4cm]{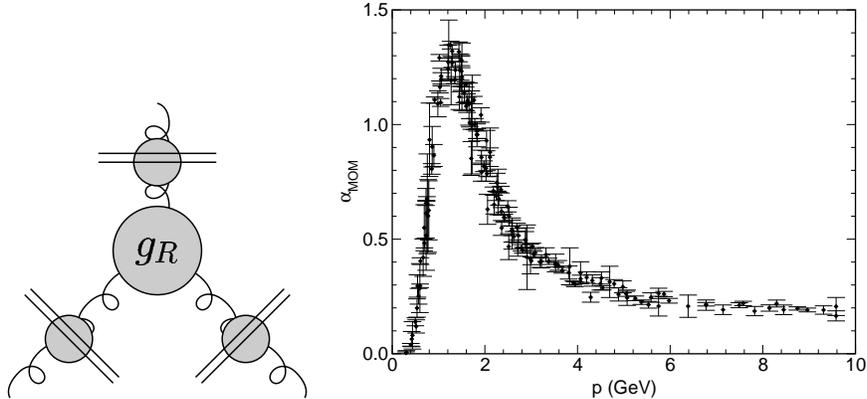} & \includegraphics[width=7cm]{alpha.eps}
\end{tabular}
\end{center}
\caption{(a) Schematic representation of the renormalization scheme used to define $\alpha_s(p)$,
the renormalized vertex is decomposed as the renormalized coupling and the three renormalized
propagators and (b) running of the coupling obtained for a wide ensemble of lattice parameters.}
\label{fig:scheme}
\end{figure}

We will present here results concerning the strong coupling constant computed 
in MOM renormalization scheme, where renormalized quantities are defined
when evaluated at the renormalization scale to take their tree level form, 
changing bare by renormalized quantities. When applied to the coupling, 
this defines its renormalized value as the amputated three gluon vertex 
(See figure \ref{fig:scheme}(a)). In terms of the scalar parts of two and three gluon 
Green functions~\cite{ope}, $G^{(2)}(p^2)$ and $G^{(3)}(p^2)$, this can be stated as:
\be\label{eq:alpha}
\alpha_{\rm MOM}(p^2) \ = \ \frac{p^6}{4 \pi}
\frac{\left(G^{(3)}(p^2)\right)^2}{\left(G^{(2)}(p^2)\right)^3}\ .
\ee
Of course, $G^{(3)}$ depends on the three entering momenta $(p_1,p_2,p_3)$. In this work we will
be limited to the simplest case, with $p_1^2=p_2^2=p_3^2$, known as symmetric MOM scheme.

The running of the coupling constant computed in the lattice  in Landau gauge and MOM
renormalization scheme has been analyzed in a series of papers~\cite{ope}, 
concluding the presence of power corrections $\sim 1/p^2$ to the purely 
perturbative running. These corrections appear in a natural way when one
performs an Operator Product Expansion (OPE) of the Green functions involved.
The OPE relates these power corrections to the existence of a gluon
condensate, $\langle A^2 \rangle$, whose value was fixed in the range $\sim 2-3\ 
{\rm GeV}^2$. This value can be understood if gluons propagate in a background
field provided by instantons~\cite{a2ins}.

\section{Instantons and low energy dynamics}

With this ingredients, the running of the coupling constant for momenta  $\gtrsim 2.5 {\rm GeV}$
can be precisely described. Nevertheless, the perturbative description, even with OPE
corrections, becomes meaningless for smaller momenta, and in particular beyond the lump (See
figure \ref{fig:scheme}(b)). Understanding this unexpected behavior of the renormalized coupling
is a key point~\cite{shi} for a full description of low energy QCD dynamics, fundamental for
problems like confinement, etc.

Yang-Mills equations  have classical solutions like instantons~\cite{bpst}, related to the
existence of zero modes in Dirac equation, and explain, for example, the difference between $\eta$
and $\eta'$ masses through the De Witten Veneziano formula~\cite{shuryak}. They could also give
the key to understand the confinement~\cite{negele}. In Landau gauge, gluon fields for an
instanton centered at the origin can be expressed as:
\be
A_\mu^a(x)\ =\ \frac{1}{g} \frac{2\overline{\eta}^a_{\mu\nu} x_\nu \rho^2}{x^2(x^2+\rho^2)}\ ,
\label{eq:amu}
\ee
with $\overline{\eta}^a_{\mu\nu}$ t'Hooft's tensor and $\rho$ a parameter that gives instanton
size. Instanton liquid models~\cite{shuryak} (ILM) in general treat with ensembles of these
structures placed randomly in space with a density $n$. In principle, in an ILM, there will be 
correlations between instantons modifying the form (\ref{eq:amu}). The effect of having a
different x-dependence is crucial when studying gluon Green functions~\footnote{Also the effect of
having instantons with different sizes is important for those cases.}, and has already 
been treated extensively~\cite{instantons}, but there are some results that do not depend on this
spatial ``profile'' of instantons.

\begin{figure}
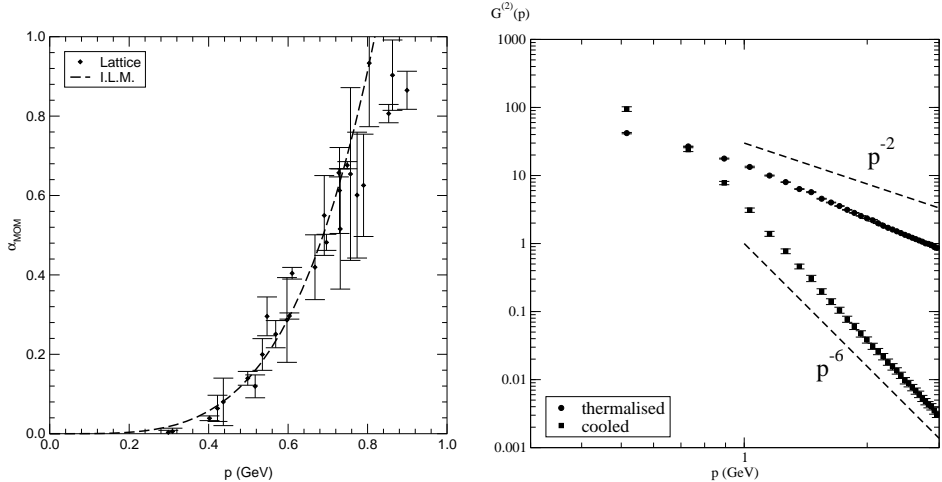

\begin{center}
\begin{tabular}{c c}
\includegraphics[width=6cm]{alpha_IR.eps} & \includegraphics[width=6cm]{propag_cool.eps}
\end{tabular}
\end{center}
\caption{(a)  and (b) gluon propagator for the original (thermalised) configurations and for the
cooled one; clearly the effect of cooling makes the propagator to fall at high energies like that
of an ILM more that that of a massless free field.}
\label{fig:IR}
\end{figure}

If we assume that long distance (low energy) correlations are dominated by instantons, gluon
Green Functions should be described by the equivalent Green functions in the field
(\ref{eq:amu}). In this independent pseudo-particle approach (called sum-ansatz in the
literature), the combination of Green functions (\ref{eq:alpha}) does depend only on the
instanton density, $n$, and not on their spatial profile or size. The prediction of an 
ILM for the renormalized coupling is then:
\be
\alpha_s^{Ins}(p^2)\ =\ \frac{1}{18\pi n} p^4\ .
\ee
This formula describes nicely the results obtained from the lattice (combining very different
lattice spacings and volumes) in the deep IR, below $1\ {\rm GeV}$ (see plot \ref{fig:IR}(a)).
From a fit to the lattice results~\cite{instantons} one can extract a value of instanton density
of around $5-7\ {\rm fm}^{-4}$. The same kind of analysis can be made separately for the Green
functions\cite{instantons} that, in this case, do depend on the profile of instantons. The
conclusions are nevertheless the same: for low energies (say below $1\ {\rm GeV}$) the running of
gluon Green functions can be understood with a simple semiclassical model. This could be related
to the confinement, as for very small energies, quantum effects over correlation functions
disappears and only classical effects survive.

\section{Green functions after cooling}

As a further crosscheck, we make use of an extended method to study instanton physics called
cooling~\cite{teper} consisting in a local iterative procedure that progressively eliminates
quantum fluctuations, leading to a smooth field distribution, where one can locate instanton
lumps. This method has nevertheless a number of known biases as instanton annihilation,
modifications of instanton size, etc. that makes hard a quantitative study. Instead
of it, we can use cooled configurations to check the results obtained for the thermalised
configurations, for example, via the calculation of gluon green functions after cooling.

Although gluon propagator in an ILM depend on the profile of instantons, its high energy limit
does not, as it is fixed by boundary conditions to:
\be
G^{(2)}(p^2) \to \frac{1}{g^2} \frac{32 n}{p^6}\ ,
\ee
while, for instance, a boson propagator behaves like $1/p^2$ (excepting logarithms). In lattice
simulations, when one applies cooling methods, gluon propagator passes from a behavior $1/p^2$ in
the thermalised configuration to $1/p^6$, as predicted by instanton models. This is a clear
indication that after eliminating short range correlations all is given by the classical
solutions. The same can be stated for three gluon vertex~\cite{instantons}, and therefore for
the  combination of Green functions that we use to define the coupling~\footnote{Of course in
cooled  configurations we can not properly speak of a ``coupling constant'' as it has no
physical meaning.} that grows like $p^4$ in all the range of energies for cooled configurations.

\section{Conclusions}

In conclusion, results coming both from the thermalised and cooled configurations confirm the
proposed image of the dominance in the large distance regime of QCD of instantons. Altogether
with the OPE description of gluon propagator and coupling, this leads to an encouraging image of
the running of the coupling where only the lump (where both classical and quantum effects are
important) lacks of explanation.

\section*{Acknowledgments}
This work has been possible thanks to financial support from French IN2P3 and Huelva University
(Spain).


\section*{References}
\begin{thebibliography}{99}


\bibitem{pdg} S.~Eidelman {\it et al.} [Particle Data Group],
\Journal{\PLB}{592}{1}{2004}. 


\bibitem{ope} P.~Boucaud, F.~De Soto, A.~Le Yaouanc, J.~P.~Leroy, 
J.~Micheli, O.~Pene and J.~Rodriguez-Quintero,
\Journal{\NPPS}{114}{117}{2003}, \Journal{\PRD}{63}{114003}{2001}.

\bibitem{shi} D.~Shirkov, \Journal{Theor. Math. Phys.}{132}{1309}{2002}


\bibitem{bpst} A.A.~Belavin, A.M.~Polyakov, A.S.~Shvarts and Y.S.~Tyupkin, 
\Journal{\PLB}{59}{85}{1975}.


\bibitem{shuryak} T.~Schafer and E.~V.~Shuryak, \Journal{\RMP}{70}{323}{1998}.

\bibitem{negele} J.~Negele \Journal{\PRD}{69}{074009}{2004}.

\bibitem{instantons} P.~Boucaud, F.~De Soto, A.~Le Yaouanc, J.~P.~Leroy, 
J.~Micheli, O.~Pene and J.~Rodriguez-Quintero, \Journal{\JHEP}{0304}{005}{2003},
\Journal{\PRD}{70}{114503}{2004} and \Journal{\JHEP}{0503}{046}{2005}.


\bibitem{a2ins}P. Boucaud, F. De Soto, A. Donini, J.P. Leroy, A. Le Yaouanc, J.
Micheli, H. Moutarde, O. Pene, J. Rodriguez-Quintero,
\Journal{\PRD}{66}{0343504}{2002}, \Journal{\NPPS}{119}{694}{2003}.

\bibitem{teper} M.~Teper, \Journal{\PLB}{162}{357}{1985}.

\end{thebibliography}
\end{document}